\documentclass[twocolumn]{aa} % for a paper on 1 column  

\usepackage{graphicx}
\usepackage{txfonts}
\newcommand{\fr}{$\nu$\ }
\newcommand{\frdot}{$\dot \nu$\ }

\begin{document}

   \title{The 1969 Glitch in the Crab Pulsar Revisited}

   \author{M. Vivekanand}

   \institute{No. $24$, NTI Layout $1$\textsuperscript{st} Stage, 
              $3$\textsuperscript{rd} Main, $1$\textsuperscript{st} 
              Cross, Nagasettyhalli, Bangalore $560094$, India. \\
              \email{viv.maddali@gmail.com}
             }

   \date{Received ---; accepted ---}

% \abstract{}{}{}{}{} 
% 5 {} token are mandatory
 
  \abstract
   { 
     Glitches are important to understand the internal structure of neutron stars. They are 
     studied using timing observations. The best studied neutron star in this respect is 
     the Crab Pulsar. The first glitch recorded in this pulsar occurred in $1969$ Sep, at an 
     epoch when timing observations (and their analysis) were still in their infancy, the 
     regularity of the observations was relatively poor, and errors on the observations were 
     relatively high in the initial stages of the observations. \cite{Lyne1993} analyzed most 
     of the available data using modern techniques, and showed that this was a typical glitch 
     of the Crab pulsar, with typical glitch parameters.
   }
   {
     This work analyses all available data, and shows that the $1969$ event in the Crab pulsar 
     is amenable to radically different interpretations.
   }
   {
     The Crab pulsar was timed by five different groups during this epoch, one at radio and the
     rest at optical frequencies. These data are available in the public domain, and have been
     analyzed using the TEMPO2 software.
   } 
   {
     The $1969$ event in the Crab pulsar can be better modeled as a typical glitch that was 
     interrupted by a (recently proposed) non-glitch speed-up event. This work also confirms
     the existence of a quasi-sinusoidal oscillation in the timing noise of the Crab pulsar,
     that was reported by \cite{Richards1970}, but with a smaller period, and with its
     amplitude and period decreasing with time, like a chirp signal. Such a coherent 
     oscillation has not been noticed so far in either the Crab or any other pulsar.
   }
   {
     This work provides an explanation for the post-glitch behavior of the Crab pulsar 
     glitches of $1969$ Sep and $2004$ Nov, and similar glitches in other pulsars, in terms 
     of the recently proposed non-glitch speed-up event. If true, then non-glitch speed-up 
     events may not be as rare as believed earlier. This work argues that it is unlikely 
     that the frequency and amplitude modulated sinusoidal variation in the timing noise is 
     due to unmodeled planetary companions.
   }


   \maketitle
%
%________________________________________________________________

\section{Introduction}

The Crab Pulsar's long term rotation is well represented by a rotation frequency $\nu$ 
and its first two time derivatives $\dot \nu$ and $\ddot \nu$; $\dot \nu$ is negative
indicating slowdown, and $\ddot \nu$ is positive, indicating decreasing slowdown over
time. Superimposed over this are two perturbations: (1) abrupt increase in the 
magnitude of \fr and \frdot roughly once in a couple of years, known as glitches, and 
(2) much slower and weaker and random variation in \fr and \frdot occurring over 
days, months and years, known as timing noise. Glitches are characterized by a sudden 
increase in rotation frequency ($\Delta \nu$, positive value) and decrease in its 
derivative ($\Delta \dot \nu$, negative value) at the epoch of the glitch $t_0$. Both 
parameters are further broken up into a change that is permanent post-glitch ($\Delta 
\nu_p$ and $\Delta \dot \nu_p$), and the remaining that decays exponentially with a 
typical timescale $\tau$ of $\approx 10$ days. Some glitches in the Crab pulsar have 
multiple decaying components, with decay time scales ranging from $0.1$ to $320$ days, 
which are not relevant here. Further details can be found in \cite{Lyne1993}, 
\cite{Espinoza2011}, \cite{Lyne2015} and \cite{Vivekanand2015}.

\begin{figure}[h]
\centering
\includegraphics[width=\hsize]{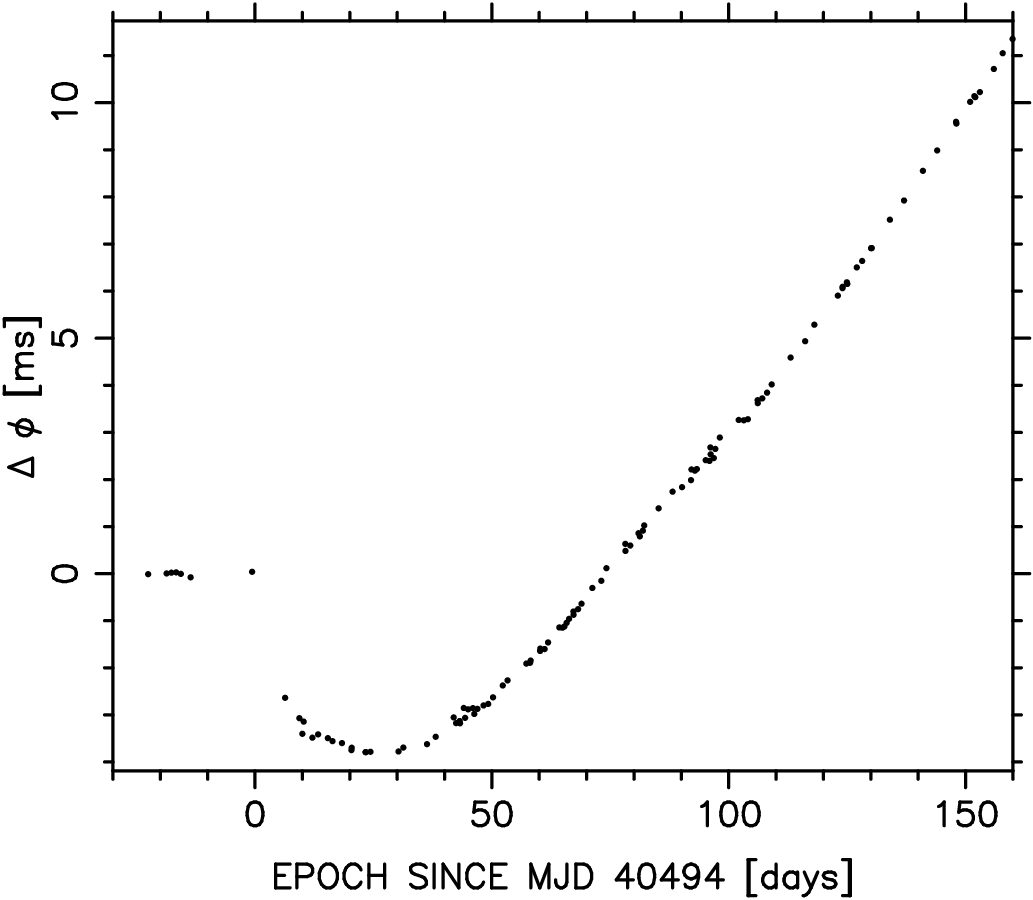}
\centering
\caption{
         Arrival Time residuals $\Delta \phi$ in milli seconds (ms) as a function of 
         time since glitch epoch $t_0 = 40494$ (MJD) for the Crab pulsar glitch of 
         $1969$. Data from the Arecibo, Princeton and Hamburg observatories have 
         been used (see Table~\ref{tbl1}). The pre-glitch reference timing model 
         of \cite{Lyne1993} has been used.
        }
\label{fig1}
\end{figure}

Glitches in the Crab pulsar are studied by first measuring the time of arrival at 
the observatory, of a fiducial point in its integrated profile, which is usually
the first of its two peaks. This is done as frequently as possible, over a 
duration of several hundreds of days enclosing the glitch. Ideally the cadence 
of the observations should be once a day, but this is rarely feasible. For 
example, even if all else was conducive, the optical observations will necessarily 
have a gap during the months when the Crab pulsar rises above the horizon during 
daylight. These site arrival times are then converted to barycenter arrival times, 
using the TEMPO2 software (\cite{Hobbs2006}). Next, one obtains the so called 
pre-glitch reference timing model, which consists of the best fit values of the 
rotation frequency and its derivatives at the glitch epoch $t_0$, labeled as 
$\nu_0$, $\dot \nu_0$ and $\ddot \nu_0$; this is done by a least squares fit to 
the pre-glitch barycenter arrival times in TEMPO2. Ideally this pre-glitch duration 
would be sufficiently long. Then the difference between the barycenter arrival 
times and the pre-glitch reference timing model, known as arrival time residuals 
$\Delta \phi$ (in seconds of time) are fit to a model of the glitch that has the 
parameters $\Delta \nu_p$, $\Delta \dot \nu_p$, $\tau$, etc. For details see 
equation $1$ of \cite{Shemar1996} and equation $1$ of \cite{Vivekanand2015}.

Figure~\ref{fig1} is a plot of $\Delta \phi$ against the epoch of observation for
the Crab pulsar glitch of $1969$, using exactly the same data, as well as exactly 
the same pre-glitch reference timing model, as that of \cite{Lyne1993}. This is 
identical to panel $1$ of their Figure $9$ in minute detail. Figure~\ref{fig1} 
clearly demonstrates the parabolic variation of post-glitch $\Delta \phi$ for a 
typical glitch (see \cite{Espinoza2011} for more examples). The glitch parameters 
derived by \cite{Lyne1993} are given in their Table $4$, and also in Table $3$ 
of \cite{Wong2001} in a modified form.

The following sections show that the arrival time residuals $\Delta \phi$ of the 
Crab pulsar during the $1969$ glitch can behave radically differently, depending
upon the choice of the pre-glitch reference timing model.

\section{Observations}

\begin{table}[h]
\begin{center}
\caption{List of the groups that timed the Crab pulsar during the $1969$ glitch, and 
the frequency of observation. The references are: ($1$) \cite{Gullahorn1977}; ($2$) 
\cite{Groth1975}; ($3$) \cite{Lohsen1981}; ($4$) \cite{Papaliolios1971}; ($5$) \cite{Duthie1971}.
} \label{tbl1}
\begin{tabular}{|l|l|l|}
\hline
 & \hfill TELESCOPE \hfill\hfill   & \hfill FREQ \hfill\hfill  \\
\hline
1 & Arecibo Observatory ($430$ MHz)& Radio \\
\hline
2 & Princeton University Observatory & Optical \\
\hline
3 & Hamburg Observatory & Optical \\
\hline
4 & Harvard Oak Ridge Observatory & Optical \\
\hline
5 & Rochester Mees Observatory & Optical \\
\hline
\end{tabular}
\end{center}
\end{table}

Table~\ref{tbl1} lists the five groups that timed the Crab pulsar glitch of $1969$.
A sixth group from the Lick Observatory (\cite{Nelson1970}) also timed the glitch,
but their published arrival times are incorrect (\cite{Horowitz1971}). 
\cite{Rankin1971} provide $25$ site arrival times, measured at different radio 
frequencies at the Arecibo observatory. These were some of the very earliest 
observations and had very large measurement errors; combining them with the rest
of the data degraded the results. Most of the arrival time data come from the 
Arecibo and Princeton observatories -- $83$ and $75$ arrival times, respectively, 
over a duration of $\approx 370$ days enclosing the $1969$ glitch, that ends at 
MJD $\approx 40660$. In the same duration the Hamburg, Harvard and Rochester 
observatories contribute $24$, $26$ and $19$ arrival times, respectively (the 
last is discussed in greater detail in the Appendix).  \cite{Lyne1993} use data 
only from the first three observatories in Table~\ref{tbl1}.

The data cadence is very non-uniform. The Hamburg and Rochester observatories do 
not contribute at all to the pre-glitch segment; the Arecibo, Princeton and 
Harvard groups contribute $44$, $25$, and $4$ arrival times, respectively, to 
this segment. The total duration of observation for each of the five 
observatories in Table~\ref{tbl1} (up to MJD $\approx 40660$) are $308$, $368$, 
$151$, $357$ and $123$ days respectively. So the average cadence of the Arecibo 
data is once in $308/83 \approx3.7$ days; for the Rochester data it is once in
$123/19 \approx 6.5$ days. The average cadence for all observatories together is 
once in $370 / (83 + 75 + 24 + 26 + 19) \approx 1.6$ days; this still falls 
short of the ideally required once a day cadence. Moreover there are large gaps in
the epochs of observations of the combined data; there are $4$ gaps in the data 
that are  larger than $11$ days, and some of the gaps larger than $5$ days are 
at the turning point of the curve in Figure~\ref{fig1}. It is important to have 
high cadence of observations at this turning point, and the last two 
observatories in Table~\ref{tbl1} contribute arrival times near this point. In
summary, this work uses $25$\% more data than \cite{Lyne1993}, the additional 
data providing greater cadence at the critical epochs in Figure~\ref{fig1}.

There is a lot of detail in the extracting and processing of this data, which 
is explained in the Appendix.

\section{Pre-glitch reference timing models}

\begin{table}[h]
\begin{center}
\caption{Parameters of five pre-glitch reference timing models, using the number 
of arrival times given in the last column; errors in the last digit are given in 
brackets.  The last two models use the pre-glitch epoch range used by 
\cite{Lyne1993}; the last model also uses the exact data used by \cite{Lyne1993}.
} \label{tbl2}
\begin{tabular}{|l|l|l|l|l|}
\hline
 & \hfill $\nu_0$ (Hz) \hfill\hfill & \hfill $\dot \nu_0 \times 10^{-10}$ (s$^{-2}$) \hfill\hfill & \hfill $\ddot \nu_0 \times 10^{-20}$ (s$^{-3}$) \hfill\hfill & \hfill N \hfill\hfill \\
\hline
1 & $30.208963495$($2$) & $-3.857040$($6$) & $1.131$($7$) & $73$ \\
\hline
2 & $30.208963503$($3$) & $-3.85698$($1$) & $1.30$($3$) & $37$ \\
\hline
3 & $30.208963503$($3$) & $-3.85704$($3$) & $0.8$($1$) & $18$ \\
\hline
4 & $30.20896350$($1$) & $-3.8571$($3$) & $-0.06 \pm 2.7$ & $10$ \\
\hline
5 & $30.20896348$($1$) & $-3.8577$($2$) & $-4.8 \pm 1.4$ & $7$ \\
\hline
\end{tabular}
\end{center}
\end{table}

\begin{figure}[h]
\centering
\includegraphics[width=\hsize]{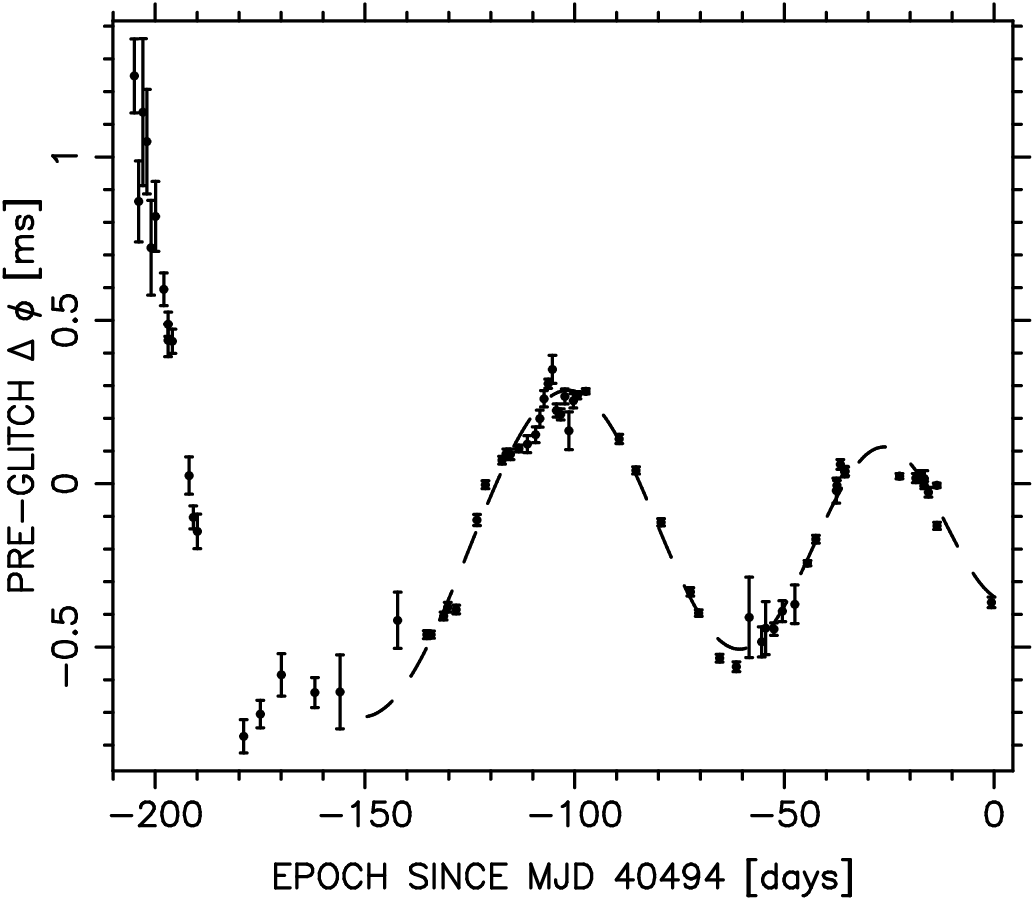}
\centering
\caption{
         Pre-glitch residuals $\Delta \phi$ as a function of 
         time since glitch epoch $t_0$ for the Crab pulsar glitch of 
         $1969$. Data from all five observatories of Table~\ref{tbl1} have 
         been used. The pre-glitch reference timing model in the first row of
         Table~\ref{tbl2} has been used. The dashed sinusoid is explained in the 
         text.
        }
\label{fig2}
\end{figure}

Table~\ref{tbl2} gives five pre-glitch reference timing models at the glitch epoch $t_0 =$
MJD $40494$, using five different ranges of pre-glitch epochs. The first model uses all 
pre-glitch data ($73$ barycenter arrival times), from epoch $\approx t_0 - 210$ to epoch 
$t_0$ days; the residuals $\Delta \phi$ with respect to this model are shown in 
Figure~\ref{fig2}. The quasi-sinusoidal variation of $\Delta \phi$ in this figure is
timing noise.

To test the stability of the results of the following sections, and also to approach 
progressively the pre-glitch reference timing model used by \cite{Lyne1993}, three
additional models were obtained, given in rows $2$ through $4$ of Table~\ref{tbl2}.
Each of them uses approximately half of the data of the previous model, by using a
pre-glitch epoch range whose starting point is roughly half of the range of abscissa 
of the previous model, but whose ending point is $t_0$. The epoch range of model $4$ 
is one month, starting at $t_0 - 30$ days and ending at $t_0$; this is exactly the 
range used by \cite{Lyne1993}. However model $4$ has $10$ data points, which is 
three more than that of \cite{Lyne1993}, because of the three additional arrival 
times from the Harvard group. Model $5$ is the same as model $4$ except that the
data is exactly that used by \cite{Lyne1993}.

The next section shows how the choice of the pre-glitch reference timing model
critically impacts upon the behavior of the post-glitch residuals. \cite{Lyne1993}
did not explore the first four pre-glitch solutions of Table~\ref{tbl2}, and
section $4.3$ discusses their solution in greater detail.

\section{Post-glitch residuals}

Figures \ref{fig3} and \ref{fig4} show the post-glitch residuals $\Delta \phi$
obtained by using the pre-glitch reference timing model in the first row of
Table~\ref{tbl2}. The data in both figures is the same. However, it is modeled 
differently as discussed below.

\subsection{A glitch interrupted by a non-glitch speed-up event}

\begin{figure}[h]
\centering
\includegraphics[width=\hsize]{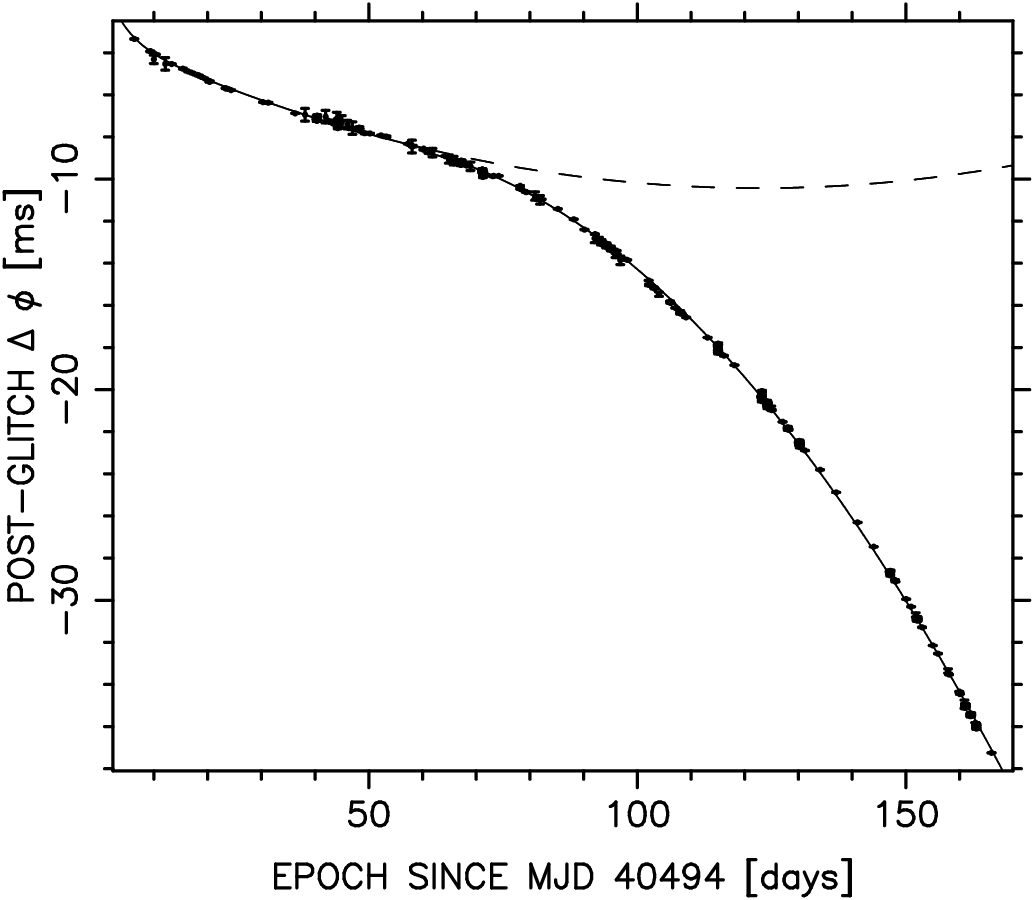}
\centering
\caption{Post-glitch residuals $\Delta \phi$ as a function of time since glitch 
         epoch $t_0$ for the Crab pulsar glitch of $1969$. Data from all five 
         observatories of Table~\ref{tbl1} have been used. The pre-glitch reference 
         timing model in the first row of Table~\ref{tbl2} has been used. The solid 
         line is the best fit curve that models the data as a typical glitch interrupted 
         by a non-glitch speed-up event occurring at about $\approx t_0 + 61$ days. The data 
         should have followed the dashed line if the glitch was not interrupted.
        }
\label{fig3}
\end{figure}

The $\Delta \phi$ of Figure~\ref{fig3} can be modeled as a typical glitch 
that is interrupted by a non-glitch speed-up event occurring at about $61$ 
days after $t_0$. The post-glitch $\Delta \phi$ from $t_0$ to $t_0 + 61$ days 
in Figure~\ref{fig3} are fit to equation $1$ of \cite{Vivekanand2015}, which 
models a typical glitch. The solid line from $t_0$ to $t_0 + 61$ days in 
Figure~\ref{fig3} represents the glitch, whose continuation is the dashed line.
The best fit glitch parameters are: $\Delta \nu_p = +0.42(1) \times 10^{-7}$ Hz, 
$\Delta \dot \nu_p = -3.9(3) \times 10^{-15}$ s$^{-2}$, $\Delta \nu_n = 2.92(5) 
\times 10^{-7}$ Hz, and $\tau = 3.7(2)$ days. These are very different from 
the values derived by \cite{Lyne1993} and given in \cite{Wong2001} (their 
$\tau = 18.7 \pm 1.6$ days).

For an uninterrupted glitch one would have expected the subsequent data to 
continue further along the dashed line in Figure~\ref{fig3}. Instead of that, 
it turns downwards in a parabolic manner, which is the hallmark of the
recently proposed  non-glitch speed-up event (\cite{Vivekanand2017}). The data 
in Figure~\ref{fig3} beyond $t_0 + 61$ days is fit to a parabolic curve which 
implies that the $\Delta \dot \nu_p$ has changed from the glitch value of 
$-3.9(3) \times 10^{-15}$ to $15.74(6) \times 10^{-15}$ s$^{-2}$. Thus
the Crab pulsar can be modeled as having undergone, around epoch $t_0 + 61$ 
days,  a persistent change of $15.74(6) \times 10^{-15}$ s$^{-2}$ in 
$\dot \nu$, which is of similar magnitude as that reported in 
\cite{Vivekanand2017}.

The above combination of two separate solutions occurs for a range of choice of 
the point of inflection $t_0 + 61$ days, from $t_0 + 42$ to $t_0 + 75$ days. The
point chosen here minimizes the discontinuity between the two solutions, which
is $0.17$ ms in Figure~\ref{fig3}. Clearly this is a radically different 
interpretation of the $1969$ event in comparison to that of \cite{Lyne1993}.
The next section discusses yet another radically different interpretation.

\subsection{An interesting alternate analysis}

This section describes an interesting analysis of the data in Figure~\ref{fig3}
which mimics some behavior of an anti-glitch \citep{Archibald2013}. This section 
is meant to show the variety in the interpretation of the data in 
Figure~\ref{fig3}, and does not imply that an anti-glitch has occurred.

\begin{figure}[h]
\centering
\includegraphics[width=\hsize]{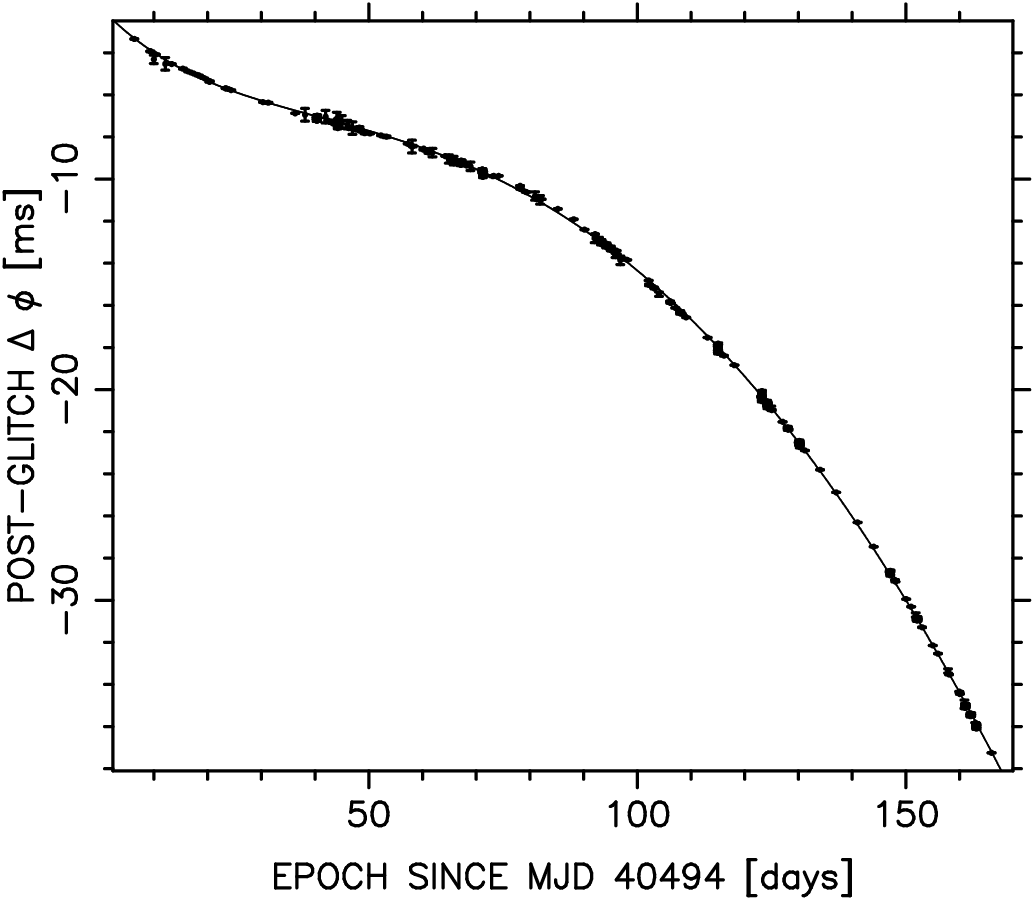}
\centering
\caption{Same data as in Figure~\ref{fig3} but now modeled as a conventional
         glitch, whose parameters are anything but conventional.
        }
\label{fig4}
\end{figure}

All post-glitch $\Delta \phi$ are fit to equation $1$ of \cite{Vivekanand2015},
which models a typical glitch. The values of the glitch parameters in
Figure~\ref{fig4} are: $\Delta \nu_p = -0.87(2) \times 10^{-7}$ Hz, $\Delta 
\dot \nu_p = +17.9(2) \times 10^{-15}$ s$^{-2}$, $\Delta \nu_n = +1.81(5) 
\times 10^{-7}$ Hz, and $\tau = 31(1)$ days, where $\Delta \nu_n$ is the 
exponentially decaying component. These are very different from the values 
derived by \cite{Lyne1993} and given in \cite{Wong2001}, where $\Delta \nu_p 
= +0.5(1) \times 10^{-7}$ Hz and $\Delta \dot \nu_p = -1.4(4) \times 10^{-15}$ 
s$^{-2}$. Numerically their values are much smaller; but more importantly 
their signs are inverted. Thus, whereas for a typical glitch in 
the Crab pulsar $\Delta \nu_p$ increases and $\Delta \dot \nu_p$ decreases 
at the glitch epoch, in Figure~\ref{fig4} it is the exact opposite, which is 
one possible property of an anti-glitch.  This is why $\Delta \phi$ in 
Figure~\ref{fig4} curves in the opposite sense to that in Figure~\ref{fig1},
beyond epoch $\approx t_0 + 30$ days. Their $\Delta \nu_n = +0.7(1) \times 
10^{-7}$ Hz, which is smaller than the value derived here, but of the same 
sign, which is why $\Delta \phi$ in Figure~\ref{fig4} curves downwards at
the glitch epoch, as in a typical glitch. Thus this is a complex behavior.

For a wide range of initial values of the parameters, the fit converges to 
the kind of curve shown in Figure~\ref{fig4}, although the final parameter 
values differ. It is reiterated once again that the $\Delta \phi$ in 
Figure~\ref{fig4} is fit to a curve that is expected from a typical glitch. 
The resulting glitch parameters, however, are quite different from those 
of a typical glitch, and appear to have some properties of an anti-glitch.

But for the discontinuity of $0.17$ ms in Figure~\ref{fig3}, the solid curves
in Figures $3$ and $4$ look remarkably similar, implying that the models in both 
figures may be equally credible, although their interpretations would be quite
different, and both interpretations being radically different from that of 
\cite{Lyne1993}.

\subsection{Importance of the pre-glitch reference timing model}

Choosing the pre-glitch reference timing model given in the first row of 
Table~\ref{tbl2} creates the dramatic difference between Figure~\ref{fig1} and 
Figure~\ref{fig3} (and equivalently Figure~\ref{fig4}). The same is true for 
the models in rows $3$ through $5$ in Table~\ref{tbl2}. However, the model in 
the second row equally dramatically reverts the behavior of $\Delta \phi$ back 
to like that seen in  Figure~\ref{fig1}. One common reason for such 
unstable behavior is lack of sufficient cadence in the timing observations,
particularly near the turning point in Figure~\ref{fig1} or near the point of 
inflection in Figures $3$ and $4$. However four out of five models result in
behavior that is similar to that seen in Figure~\ref{fig3}. It is therefore
argued that this behavior is more credible than that of \cite{Lyne1993}.

Using the pre-glitch reference timing model in the last row of Table~\ref{tbl2} 
causes $\Delta \phi$ to behave as in Figure~\ref{fig3}, and not as in 
Figure~\ref{fig1}, although this model was derived using exactly the same data 
that was used by \cite{Lyne1993}. Their values for the model parameters are
$\nu_0 = 30.20896350835(1)$ Hz and $\dot \nu_0 = -3.856965(5) \times 10^{-10}$ 
s$^{-2}$. These differ from the values of Table~\ref{tbl2} by 2.8 and 3.7 
standard deviations respectively. They do not state the value of $\ddot \nu_0$ 
used; so it can either be $0$ (i.e., they may not have used this parameter at 
all in TEMPO2), or a value that is consistent with the Crab pulsar's braking 
index of $2.51(1)$ (\cite{Lyne1993}), which is $1.236 \times 10^{-20}$ s$^{-3}$;  
Figure~\ref{fig1} has been obtained using this value of $\ddot \nu_0$, which
is very different from the corresponding value in Table~\ref{tbl2}. Thus the
glitch parameters of \cite{Lyne1993} are significantly different from those 
in row $5$ of Table~\ref{tbl2}. Their values do not correspond to a minimum 
$\chi^2$ solution of TEMPO2; if one starts with their values, TEMPO2 converges 
to the values of model $5$ in Table~\ref{tbl2}.

The practice of deriving only $\nu_0$ and $\dot \nu_0$, and using a $\ddot 
\nu_0$ that is consistent with the Crab pulsar's braking index, merits 
scrutiny. Braking index defines the secular (long term) rotation history 
of any pulsar. The above practice is justified if the pre-glitch reference 
timing model is obtained over several hundreds of days, or equivalently 
several cycles of timing noise (as has been done for the best three Crab 
glitches in \cite{Lyne2015}). By obtaining $\nu_0$ and $\dot \nu_0$ over a
month of data span, one is fitting into the local timing noise curve, 
whose variations (which are quite severe in the Crab pulsar) can lead to 
even negative values of $\ddot \nu_0$. This is evident in the several 
negative $\ddot \nu_0$ values listed in the Jodrell Bank Crab Pulsar 
Monthly Ephemeris\footnote{http://www.jb.man.ac.uk/pulsar/crab.html} 
(CGRO format) (\cite{Lyne1993}).

In summary, the pre-glitch reference timing model in the last row of 
Table~\ref{tbl2} shows that same post-glitch behavior as in Figure~\ref{fig3}, 
while the corresponding solution of \cite{Lyne1993} shows the behavior in 
Figure~\ref{fig1}; these are dramatically different behavior at epochs beyond
$\approx 50$ days after the glitch. Numerically the two solutions are
statistically different, although they are both supposed to be derived from
exactly the same data.

\section{Periodic variation of timing noise}

In Figure~\ref{fig2} the $\Delta \phi$ variation is known as timing noise. It has
been known that the timing noise of the Crab pulsar has quasi-periodic variation
of periods ranging from $200$ to $800$ days (\cite{Lyne1993}). \cite{Richards1970}
have reported a quasi-periodic variation of a much smaller period, of $77 \pm 7$ 
days, between May $10$ and Sep $16$ of $1969$. The variation seen from epoch $t_0 
- 150$ to $t_0$ days (almost the same range as that of \cite{Richards1970}) looks 
like a frequency modulated sine wave, with a slightly smaller period.  It can be 
modeled as a sinusoid whose period and amplitude are decreasing with time. The 
dashed curve in Figure~\ref{fig2} represents the best fit sinusoid having the 
formula
\begin{equation}
\Delta \phi(t) = a + (b + qt) \times \sin \left ( \frac{2.0 \pi t}{c + pt} 
                 + d \right )
\end{equation}
\noindent where $t$ is the epoch in days (with respect to $t_0$ in 
Figure~\ref{fig2}). At epoch $t = 0$, the period of the sinusoid is $c = 
55.7 \pm 2.1$ days, and its amplitude is $b = 0.21 \pm 0.02$ ms. The 
period decreases at the rate $p = -0.15 \pm 0.01$ days per day of epoch,
while the amplitude decreases by $q = -0.002 \pm 0.0002$ ms per day.  
This is reminiscent of a non-linear chirp signal of a radar. To the best 
of our knowledge such a signal has so far not been reported in the timing 
noise of the Crab or any other pulsar. The $\Delta \phi$ variation before 
epoch $t_0 - 150$ may lead one to speculate whether this is merely the 
tail end of a much larger chirp signal; however this data does not fit 
the later data as a coherent oscillation.

\section{Discussion}

This work demonstrates that the $1969$ event in the Crab pulsar can be
better understood as a typical glitch that was interrupted by a non-glitch 
speed-up event \citep{Vivekanand2015}. This is based on the variation of 
post-glitch $\Delta \phi$ as a function of epoch shown in Figure~\ref{fig3}.
These results hold even after ignoring the nine radio data that have flag
''A'' and one data that has flag ''C'', which have higher error bars.

Although this is the first time that such behavior has been explicitly 
highlighted in any pulsar, it appears that this has been seen before but 
not recognized as such. The Crab pulsar glitch of $2004$ Nov (at MJD 
$53331.17$) appears to be a similar event (see panel in row $2$, column $3$ 
of Figure $7$ of \cite{Espinoza2011}), as also the event in PSR J$1740+1000$ 
at MJD $54747.6$ (row $11$, column $3$ of same figure). Similar events 
occurring very close to the glitch epoch are probably evident in PSR 
J$0631+1036$ at MJD $50183.5$ (row $3$, column $2$), in PSR B$1702-19$ at 
MJD $48902.1$ (row $8$, column $3$) and in PSR J$1847-0130$ at MJD $53426$ 
(row $16$, column $3$), in the same figure of \cite{Espinoza2011}. If this 
turns out to be correct, then non-glitch speed-up events no longer appear 
to be very rare events. In summary, the behavior reported here, for the 
$1969$ glitch in the Crab pulsar, has been seen in at least one later glitch 
of the Crab pulsar, as well as in glitches of some other pulsars; but it
was so far not recognized as being related to the recently proposed 
non-glitch speed-up event \citep{Vivekanand2017}.

\cite{Vivekanand2017} could not determine the time scale of occurrence
of the non-glitch speed-up event in the Crab pulsar, since the data used 
had a cadence of once a month. The analysis of section $4.1$ shows
that the non-glitch speed-up event sets in on a time scale of less than, 
or equal to, a month. Further, this event 
occurred $\approx 61$ days after a glitch, while that reported by 
\cite{Vivekanand2017} occurred $\approx 1200$ days after a glitch. Therefore 
non-glitch speed-up events can apparently occur at any time with respect to 
a glitch.

This work verifies the the quasi-sinusoidal variation in the timing noise 
of the Crab pulsar in later $1969$, that was reported by \cite{Richards1970}.
However, the sinusoid appears to be frequency and amplitude modulated 
(Figure~\ref{fig2}), and has been noticed for the first time in the Crab or 
any other pulsar. There are very few pulsars in which timing noise is almost 
periodic (see Figures $3$ and $13$ of \cite{Hobbs2010}), but those 
periodicities are in the hundreds of days, unlike here. This should form a
constraint for theories of timing noise, which is as yet an unexplained 
phenomenon. One of the possible explanations is unmodeled planetary 
companions (\cite{Richards1970, Cordes1993}). \cite{Thorsett1993} searched 
for Jupiter sized planets orbiting around pulsars, by looking for 
periodically modulated timing noise. \cite{Cordes2008} discuss how such 
planets can cause other effects in pulsars such as nulling and profile 
changes. \cite{Richards1970} suggested that the quasi-periodic variation
they noticed could be caused by an Earth sized planetary companion of
the Crab pulsar.  Here it will be argued that an Earth sized planetary 
companion can, in principle, explain the amplitude and periodicity 
observed in Figure~\ref{fig2}, as well as their decrease with time, but 
that such a scenario is inconsistent with the rest of the observations,
at least currently.

Assuming that the Crab pulsar had a planetary companion for the duration
of the periodic signal in Figure~\ref{fig2}, which is about $\approx 150$ 
days, its mass is given by
\begin{equation}
(M_p \sin i)^3  = 1.07 \times 10^{-3} \times (a_0 \sin i)^3 \times 
                 (M_p + M_0)^2 / T^2
\end{equation}
\noindent where the planetary mass $M_p$ and Crab pulsar's mass $M_0$ are
in solar masses, the semi-major axis of Crab pulsar's orbit is $a_0$ in
light seconds, $i$ is the angle of inclination of the orbit, and the 
orbital period of both objects is $T$ in days; this is the mass function 
of binary orbits (see equation $5$-$1$ of \cite{Manchester1977}). The 
sinusoid fit in  Figure~\ref{fig2} gives $a_0 \sin i = 0.21$ ms and 
$T = 55.7$ days at $t = 0$. Assuming the mass of the Crab pulsar to be the 
standard $1.4$ solar masses, one obtains $M_p \sin i = 0.6$ Earth masses. 
Assuming that the inclinations angle $i$ is not an extreme value, one 
can conclude that an Earth sized planetary companion can cause the 
periodicity of $\approx 56$ days while orbiting the Crab pulsar. Since
the eccentricity of the orbit is not known, the minimum distance of 
approach of the planetary companion to the Crab pulsar is unknown.
The decreasing period and amplitude of the sinusoid with time are 
consistent with a planetary companion falling into the Crab pulsar, due 
to orbital decay.

The first problem with this scenario is how did the planetary companion 
appear suddenly orbiting the Crab pulsar. The rapidly decreasing $\Delta 
\phi$ before $t_0 - 150$ in Figure~\ref{fig2} can probably be used to 
argue that it represents the capture of this planetary companion by the 
Crab pulsar.

Next, one would have to assume that this planetary companion is made up 
of neutral material, so as not to encounter intense electrical and magnetic 
resistance within the magnetosphere of the Crab pulsar. However, 
it is not clear whether material that was neutral to start with will 
continue to remain neutral within the magnetosphere of the Crab 
pulsar, which is expected to be the site of intense radiation, even in
the closed field lines of the Crab pulsar. Further, the planetary 
companion probably can not avoid being vaporized by the Crab pulsar's
radiation in the magnetosphere (see \cite{Cordes2008}). Therefore 
it is not clear that such a planetary companion can survive the Crab 
pulsar's environment and remain stable for $\approx 150$ days.

Finally, the sinusoidal variation of timing noise appears to have been 
terminated abruptly just before the $1969$ glitch. In principle the 
planetary companion could free fall into the Crab pulsar instantaneously,
but one needs a physical mechanism for the planetary companion to 
suddenly lose its entire angular momentum and free fall into the Crab
pulsar. This would be as dramatic as its sudden appearance $\approx 150$
days earlier. If such a scenario is credible, then it is difficult not to 
associate the Crab pulsar glitch of $1969$ with the impact of this 
planetary companion upon the surface of the neutron star. However, the 
Crab pulsar has undergone several glitches where such planetary 
companions have not been invoked.

\cite{Brook2014} report the possible impact of an asteroid, much smaller 
than Earth, on PSR J$0738-4042$. This caused non-periodic torque
variations, and also caused changes in the integrated profile of the
pulsar. Profile changes are expected since the planetary companion is
expected to be vaporized and ionized by the pulsar radiation, which 
would cause additional electric current on the magnetic field lines.
However the original observers (see Table~\ref{tbl1}) have not reported
any profile variations in the Crab pulsar around the epoch of the 
sinusoidal variation in Figure~\ref{fig2}. This further argues against
the possibility of an in falling planetary companion.

In summary, while it is in principle possible to explain the periodicity 
observed in Figure~\ref{fig2} as being due to an orbiting Earth sized 
planetary companion, whose orbit is decaying with time, currently such a 
scenario does not appear to be credible, unless more detailed theory is 
invoked.

\cite{Starovoit2017} have recently shown that an Earth sized planetary 
companion may be orbiting PSR B$0329+54$, but the orbital period is much
longer, $\approx 28$ years, in comparison to the $\approx$ weeks that 
are involved here.

%__________________________________________________________________

%______________________________________________________________

\begin{acknowledgements}

I thank Francis Graham-Smith and Andrew G. Lyne for useful discussion in spite of 
disputing the significance of the conclusions. I am very grateful to Joel Weisberg
for bringing to my attention the references of \cite{Richards1970} and 
\cite{Starovoit2017}, and for comments.

\end{acknowledgements}

\appendix
\section{Details of the Observations}

This work uses only a small fraction of the data that is available in the references in 
Table~\ref{tbl1}. The number of arrival times in each of the references are $615$, $348$, 
$600$, $32$ and $239$, respectively, totaling $1834$, tabulated as Julian days up to 
the eleventh decimal place, along with a timing accuracy in micro seconds ($\mu$s). In 
addition, the dispersion constant has been tabulated for the radio data (six digit 
number in units of sec.MHz$^2$), while the Gregorian dates of the arrival times (year, 
month day, hours, minutes and seconds, the last up to the sixth decimal place) have been 
tabulated for the Princeton, Hamburg and Harvard data. Thus, there are about eleven 
thousand numbers to be read out from five pdf files, about $40$\% of them having a large 
number of digits. The numbers could be copied and pasted from only the Princeton 
reference, and that too in a very limited manner. The rest of the references are scanned 
files. 

The pdf file of each observatory was input to two independent Optical Character 
Recognition (OCR) 
software\footnote{https://www.onlineocr.net/}\footnote{http://free-online-ocr.com/},
after it was broken up into much smaller pieces, whose size depended upon the quality 
of the print in that pdf file. The poor print quality of the Hamburg pdf file proved 
to be particularly challenging. The text output of each OCR software was 
visually compared with the corresponding pdf file, number by number, and corrections 
were made if required. Then the text outputs of the two OCR software were compared 
with each other using the \textit{diff} utility of the Linux operating system, and 
corrections made if required. This was done sufficiently slowly to account for human
fatigue. After about a month, the final text outputs were once again visually compared, 
number by number, with the corresponding pdf files.

Barycenter arrival times are tabulated only by the Princeton, Hamburg and Harvard
groups. These can not be combined together because each group has used a different 
method of barycentric correction. These three groups tabulate site arrival times 
in both the Gregorian and Julian dates. However, The Hamburg Julian dates 
have been derived from the corresponding Gregorian dates using a different formula 
(this is explained in their reference). Therefore their barycenter arrival times 
can not be used, even just by themselves, in a modern software such as TEMPO2. This 
occurred because of the need to reconcile the discontinuous time scale in which 
site arrival times are measured (Coordinated Universal Time or UTC) with the 
continuous Terrestrial Time scale (TT) in which barycenter arrival times are 
computed; this process was still evolving in those days, as also was the method 
of barycentric correction.

In this work, the Gregorian site arrival times of these three groups have been 
processed by the software routine \textit{iauCal2jd}, available in the IAU 
Standards of Fundamental Astronomy (SOFA) 
software library\footnote{http://www.iau-sofa.rl.ac.uk/}, to obtain the 
corresponding Julian site arrival times. Next, the Julian site arrival times of 
all groups were processed using the TEMPO2 software to obtain the barycenter 
arrival times, using the ephemeris in 
Table~\ref{tbla1}\footnote{http://www.atnf.csiro.au/people/pulsar/psrcat/}.

\begin{table}[h]
\begin{center}
\caption{Crab Pulsar's ephemeris used for barycentric correction, obtained from the 
ATNF pulsar catalog (\cite{Manchester2005}).
} \label{tbla1}
\begin{tabular}{|l|l|}
\hline
\hfill PARAMETER \hfill\hfill   & \hfill VALUE \hfill\hfill  \\
\hline
RAJ (h:m:s) & +05:34:31.973 \\
\hline
DECJ (d:m:s) & +22:00:52.06 \\
\hline
PMRA (mas/yr) & -14.7 \\
\hline
PMDEC (mas/yr) & +2.0 \\
\hline
POSEPOCH (MJD) &  40706 \\
\hline
\end{tabular}
\end{center}
\end{table}

TEMPO2 already has the geocentric coordinates of the Arecibo and Princeton observatories;
for the rest of the observatories they are given in Table~\ref{tbla2}. They are obtained
by first getting the geodetic coordinates from the observatory web sites, and also from
general sites\footnote{http://www.eso.org/~ndelmott/obs\_sites.html}. They have also been
verified using Google maps\footnote{https://www.google.co.in/}. The Rochester geodetic 
coordinates are given in their reference. Then the geodetic coordinates are converted to 
geocentric coordinates using the software routine \textit{iauGd2gc} of SOFA.

\begin{table}[h]
\begin{center}
\caption{Geocentric coordinates of three observatories in meters. Their accuracy is expected
to be $\approx 100$ meters, which is sufficient to time the Crab pulsar.
} \label{tbla2}
\begin{tabular}{|l|l|l|l|}
\hline
\hfill TELESCOPE \hfill\hfill   & \hfill X \hfill\hfill & \hfill Y \hfill\hfill & \hfill Z \hfill\hfill \\
\hline
Hamburg & 3743367 & 676245 & 5102506 \\
\hline
Harvard & 1489772 & -4467571 & 4287249 \\
\hline
Rochester & 1023510 & -4582206 & 4303547 \\
\hline
\end{tabular}
\end{center}
\end{table}

Consistency checks were done on the data extracted from each observatory file. For
the Arecibo group, the first three fits in their Table $2$ were verified. The 
derived values of $\dot \nu_0$ and $\ddot \nu_0$ were consistent with their values 
within errors. However the derived $\nu_0$ differed by about $0.001$ to $0.01$
micro Hertz ($\mu$Hz), which is probably on account of a residual phase gradient
due to differences in barycentric correction. Figure $3$ in the reference of the 
Princeton group has been verified using the values of $\nu_0$, $\dot \nu_0$ and 
$\ddot \nu_0$ in their Table $3$, although the curvatures are slightly different 
probably due to the same reasons. Similarly some figures in the reference of the 
Hamburg group have been reproduced with minor differences.

Alignment of data from different observatories is achieved by using known site 
arrival time offsets, that are listed in Table~\ref{tbla3}. It was found that 
these offsets indeed aligned the data. The offsets proposed by \cite{Horowitz1971} 
for the Harvard and Rochester data were found to be unnecessary.

\begin{table}[h]
\begin{center}
\caption{Site arrival time offsets with respect to the Princeton observatory
data. The Hamburg observatory data has two offsets, the first during 
$1969$/$1970$, and the second during $1970$/$1971$ (\cite{Lohsen1981}). The
Arecibo offset is obtained from \cite{Slowikowska2009}, and is very close to 
the value used by \cite{Groth1975}.
} \label{tbla3}
\begin{tabular}{|l|l|}
\hline
\hfill TELESCOPE \hfill\hfill   & \hfill OFFSET \hfill\hfill \\
\hline
Arecibo & -235 ($\mu$S) \\
\hline
Hamburg1 & -2430 ($\mu$S) \\
\hline
Hamburg2& +116 ($\mu$S) \\
\hline
\end{tabular}
\end{center}
\end{table}

Although the Rochester group tabulate $239$ arrival times, they have $\approx 10$
observations per day; so they have essentially observed for only $29$ independent
days.

In this work the solar system ephemeris DE200 of JPL has been used, and TEMPO2 has 
been used in in the TEMPO1 compatibility mode; this results in the barycentric 
correction being done in the TDB units. The results have been verified using the 
more modern DE421 ephemeris, and working in the TCB units.

%-------------------------------------------------------------------

\end{document}